\documentclass[aps,twocolumn,raggedbottom,showpacs,nobalancelastpage,amssymb,groupedaddress]{iopart}
\usepackage{graphicx}
\usepackage{amsmath-iop}
\usepackage{amssymb}
\usepackage[latin1]{inputenc}
\usepackage[T1]{fontenc}
\usepackage[english]{babel}
\usepackage{bm}
\usepackage{indentfirst}
\usepackage{subfigure}
\usepackage{setspace}
\usepackage{amsfonts}
\usepackage{eucal}
\usepackage{amstext}
\usepackage{exscale}
\usepackage[usenames]{color}
\usepackage{cases} 
\usepackage{stmaryrd}

\newcommand{\intg}[2]{\int_{#1}^{#2} \!\!\mathrm{d}}
\newcommand{\lt}{Laplace Transform}
\newcommand{\lts}{Laplace Transforms}
\newcommand{\eq}[1]{Eq.\;\,\!\eqref{#1}}
\newcommand{\fig}[1]{Fig.\;\,\!\ref{#1}}
\newcommand{\ii}{\mathrm{i}}

\newcommand{\Ks}{K} 
\newcommand{\atp}[2]{\genfrac{}{}{0pt}{1}{\textstyle{#1}}{\textstyle{#2}}} 

\begin{document}
\title{Dynamics of a qubit coupled to a broadened harmonic mode at finite detuning}
\author{ F.\  Nesi$^1$, M.\  Grifoni$^1$, and E.\ Paladino$^2$}
\address{$^1$ Institut f\"{u}r Theoretische Physik,
Universit\"at Regensburg, 93035 Regensburg, Germany}
\address{$^2$ MATIS INFM-CNR \& Dipartimento di Metodologie Fisiche e Chimiche,
Universit\`a di Catania, 95125 Catania, Italy}
\date{\today}
\begin{abstract}
We study the dynamics of a symmetric two-level system strongly
coupled to a broadened harmonic mode. Upon mapping the problem
onto a spin-boson model with peaked spectral density, we show how
analytic solutions can be obtained, at arbitrary detuning and
finite temperatures, in the case of large Q-factors of the
oscillator. In general {\em two} or more oscillation frequencies
of the two-level particle are observed as a consequence of the
entanglement with the oscillator. Our approximated analytical solution well
agrees with numerical predictions obtained within the
non-interacting blip approximation.
\end{abstract}


\allowdisplaybreaks

\section{Introduction}

A prominent physical model to study dissipative and decoherence
effects in quantum mechanics is the spin-boson model
\cite{Leggett87,Wei,Gri98}. Currently, we witness its revival since it
allows a quantitative description of solid-state quantum bits
(qubits) \cite{Makhlin01}. A more realistic description requires
the inclusion of external control fields  as well as of a
detector. In the spin-boson model, the environment is
characterized  by a spectral density $G(\omega)$.
 If the environment is
formed by a quantum detector which itself is damped by Ohmic
fluctuations, the form of the spectral density can become
non-trivial, as it reflects also internal resonances of the
detector. 
An example is provided by a flux-qubit read out by a dc-SQUID
\cite{Wal00,Chi03,Patrice04} whose plasma resonance at $\Omega$
gives rise to an effective spectral density $G_{\rm eff}(\omega)$
for the qubit with a peak at $\Omega$ \cite{Tian02,vdW}, cf.\  Eq.\
(\ref{Gw}) below. Recently, the coherent coupling of a single
photon mode and a superconducting charge qubit has also been
studied \cite{Wallr04}. Until now, the effects of such a
structured spectral density on the decoherence properties of a qubit have been studied in
\cite{Thorwart00,Smi03} within a perturbative approach in $G_{\rm
eff}$. It was shown in \cite{Tho03,Kle03} that such 
a perturbative scheme breaks down for strong qubit-detector
coupling, and when the qubit and detector
 frequencies are comparable.
 %
Hence, in \cite{Tho03,Goo03,Goo05} the dynamics was investigated by mapping the spin-boson problem onto the equivalent situation of a two-state system (TSS) coupled to a harmonic oscillator (HO), the latter coupled to an Ohmic bath with spectrum $G_{\rm
Ohm}$. By considering the TSS and HO as the relevant system, analytic solutions perturbative in $G_{\rm Ohm}$ were obtained.

  In this work we show how to investigate the dynamics of a spin-boson system
  with a structured
 environment, in the case of a strong coupling between qubit and
  detector. 
%
%
 We evaluate  the dynamics upon starting from the (nonperturbative in $G_{\rm eff}$) Non-Interacting Blip Approximation (NIBA) \cite{Leggett87,Wei}. Analytical results, valid also at finite detuning, are obtained by approximating the NIBA kernels up to first order in the 
 detector-bath coupling strength.
The paper is organized as follows: In the next Section we will introduce the model. 
 Then, in Sec. \ref{sniba} we discuss the well-known and widely used Non-Interacting Blip Approximation and its predictions. Analytical results for the dynamics are derived in Sec. \ref{swca}.

\section{The model}

In this work we consider the spin-boson Hamiltonian describing the interaction of a symmetric TSS with a structured environment. It reads \cite{Leggett87,Wei}
\begin{align}
\begin{split}
{H_{SB}}(t)&=-\frac{\hbar\Delta}{2} \sigma_x
+ \frac{1}{2}\sigma_z \hbar \sum_k \tilde{\lambda}_k
(\tilde{b}_k^{\dagger} +\tilde{b}_k) + \sum_k \hbar
\tilde{\omega}_k \tilde{b}_k^{\dagger} \tilde{b}_k\, ,
\label{hamtot}
\end{split}
\end{align}
where $\sigma_i$ are Pauli matrices and $\hbar\Delta$ is the tunnel splitting. 
 Moreover, 
$\tilde{b}_k$ 
is the annihilation 
operator of the $k-$th bath mode with frequency
$\tilde{\omega}_k$. 
In the spin-boson model the influence of the environment is fully characterized by a so-called spectral function, which we assume to be of the form 
\begin{align}
\!\! G_{\rm eff}(\omega)=\sum_k \tilde{\lambda}_k^2
\delta(\omega-\tilde{\omega}_k)=  \frac{2 \alpha \omega
\Omega^4}{(\Omega^2-\omega^2)^2+(\Gamma\omega)^2}.
%
%
\label{Gw}
\end{align}
%
It has a Lorentzian peak of width $\Gamma$
 at the
characteristic frequency $\Omega$, and behaves
Ohmically at low frequencies with the dimensionless coupling
strength $\alpha = \lim_{\omega \rightarrow 0} G_{\rm
eff}(\omega)/2\omega$. 
As shown in \cite{Garg85}, such spin-boson Hamiltonian can be exactly mapped onto that of a TSS coupled to a single harmonic oscillator mode of frequency $\Omega$ with coupling strength $g$. The HO iteslf interacts with a set of harmonic oscillators with spectral density of the continuous bath modes being $G_{\rm Ohm}(\omega)=\kappa \omega$. The mapping between the two models is completed with $\Gamma=2\pi\kappa\Omega$ and $\alpha=8 \kappa g^2/\Omega^2$. Notice that the oscillator can e.g. represent a dc-SQUID with plasma frequency $\Omega$ which couples inductively to a superconducting flux qubit \cite{Tian02,vdW}. The damping of the dc-SQUID is due to its coupling to an electromagnetic environment.

If the damping of the harmonic oscillator is small ($\kappa \ll 1$), as in typical experiments where the dc-SQUIDs are typically underdamped, then it would seem more convenient to use the mapping, and consider the qubit coupled to the dc-SQUID as unique quantum system. 
However, since such a system has an infinite Hilbert space, the inclusion of dissipation is typically done \cite{Tho03,Goo05} upon truncation of the system's Hilbert space to a few relevant levels (which is the case e.g. if $\hbar \Omega\,, \hbar\Delta \gg k_{\rm B}T$).
This led in \cite{Tho03} to find analytical results for the resonant case $\Delta=\Omega$ within a three-level approximation. In the present work analytical results valid at finite detuning $\Delta\ne\Omega$ are obtained by focusing  on the spin-boson model \eqref{hamtot}.
The advantage of this approach is that the reduced density matrix has rank 2. The peculiar feature of the peaked spectrum \eqref{Gw} is reflected in the form of the bath correlation functions, cf. \eqref{Q} and \eqref{Q''niba} below.

Specifically, in the spin-boson model the environmental effects are captured in the so-called bath correlation function 
\begin{align}
 Q(\tau)\equiv Q'(\tau)+\ii Q''(\tau)=\frac{1}{\pi\hbar}\intg{0}{\infty}\omega \frac{G_{\rm eff}(\omega)}{\omega^2}\left[ \coth{\left(\frac{\hbar\omega\beta}{2}\right)}(1-\cos{\omega t})+\ii \sin{\omega t} \right] ,
\end{align}
which for the effective spectral density \eqref{Gw} reads 
\begin{subequations} \label{Q}
\begin{align}
Q'(\tau)&= X \tau+L \left(e^{-\frac{\Gamma}{2}\tau}\cos{\bar{\Omega}\tau}-1\right)+ Z e^{-\frac{\Gamma}{2}\tau}\sin{\bar{\Omega}\tau}+Q'_{Mats}(\tau) \label{Q'niba}, \\
Q''(\tau)&= \pi\alpha-e^{-\frac{\Gamma}{2}\tau} \pi\alpha \left(N
\sin{\bar{\Omega}\tau}+ \cos{\bar{\Omega}\tau}\right),
\label{Q''niba}
\end{align}
\end{subequations}
being $\bar{\Omega}=\sqrt{\Omega^2-\tfrac{\Gamma^2}{4}}$ and 
\begin{align}
X&=\frac{2\pi \alpha}{\hbar\beta}, \\
L&=\frac{\pi \alpha}{\Gamma \bar{\Omega}}\: \frac{1}{\cosh{(\beta\bar{\Omega})}-\cos{\left(\beta\tfrac{\Gamma}{2}\right)}} \left[\left(\tfrac{\Gamma^2}{4}-\bar{\Omega}^2\right) \,\sinh{(\beta\bar{\Omega})}+\Gamma\bar{\Omega}\,\sin{\left(\beta\tfrac{\Gamma}{2}\right)}\right],  \\
Z&=\frac{\pi \alpha}{\Gamma \bar{\Omega}}\:
\frac{1}{\cosh{(\beta\bar{\Omega})}-\cos{\left(\beta\tfrac{\Gamma}{2}\right)}}
\left[-\Gamma\bar{\Omega}\,\sinh{(\beta\bar{\Omega})}+
\left(\tfrac{\Gamma^2}{4}-\bar{\Omega}^2\right)\,\sin{\left(\beta\tfrac{\Gamma}{2}\right)}\right],  \\
N&=\frac{1}{\Gamma \bar{\Omega}}\: \left(\tfrac{\Gamma^2}{4}-\bar{\Omega}^2\right).
\end{align}
$Q'_{Mats}(\tau)$ is a function of Matsubara frequencies and has the form 
\begin{equation} \label{Qmats}
Q'_{Mats}(\tau)=-4 \pi\alpha \frac{\Omega^4}{\hbar\beta}\sum_{n=1}^{+\infty} \dfrac{1}{(\Omega^2+\nu_n^2)^2-\Gamma^2 \nu_n^2} \left[\dfrac{e^{-\nu_n \tau}-1}{\nu_n}\right], 
\end{equation}
with the Matsubara frequencies defined as $\nu_n\equiv\dfrac{2\pi}{\hbar\beta}n$.
For temperatures high enough ($k_B T \gg
\tfrac{\hbar\Gamma}{2\pi}$), contributions coming from the Matsubara term can be neglected, as done in the rest of this work.

The qubit dynamics is described by the reduced density operator
$\rho(t)$ obtained by tracing out all environmental degrees of
freedom. We investigate the population difference
$P(t):=\langle\sigma_z\rangle_t=\tr\{\rho (t) \sigma_z\}$.
Such a dynamical quantity $P(t)$ obeys the exact generalized master equation (GME) \cite{Wei}
\begin{equation} \label{gme0}
\dot{P}(t)=-\intg{0}{t}t'  
\Ks(t-t')P(t')
, \qquad t>0.
\end{equation}
with the kernels $\Ks(t)$ being a series expression in the number of tunneling transitions.

Since Eq. \eqref{gme0} involves only convolutions, 
it can be solved by using \lts. Hence, the GME transforms 
as  
\begin{equation} \label{P}
P(\lambda)=\dfrac{1}
{\lambda+\Ks(\lambda)},
\end{equation}
where the same symbols $P(\lambda)$ and $K(\lambda)$ for the Laplace transform of $P(t)$ and
$K(\tau)$ have been used, respectively. 
From Eq.\ \eqref{P}, it follows that in order to obtain $P(t)$ it can be enough to solve the pole equation
\begin{equation} \label{peq}
\lambda+\Ks(\lambda)=0
\end{equation}
and then antitrasform back to the time space.

Due to the intricate form of the exact kernel $\Ks(t)$ (or $\Ks(\lambda)$), Eqs. \eqref{gme0} (or \eqref{P}) cannot be solved neither numerically nor analytically. 
We must therefore invoke some approximations. For the symmetric spin-boson model \eqref{hamtot}, the so-called Non-Interacting Blip Approximation (NIBA) discussed in the next Section is known to yield reliable results over the \textit{whole} regime of parameters.

\section{Non-Interacting Blip Approximation (NIBA)} \label{sniba}

\begin{figure}[htb!]
\begin{center}
\includegraphics[width=0.7\textwidth,bb=37 56 718 527]{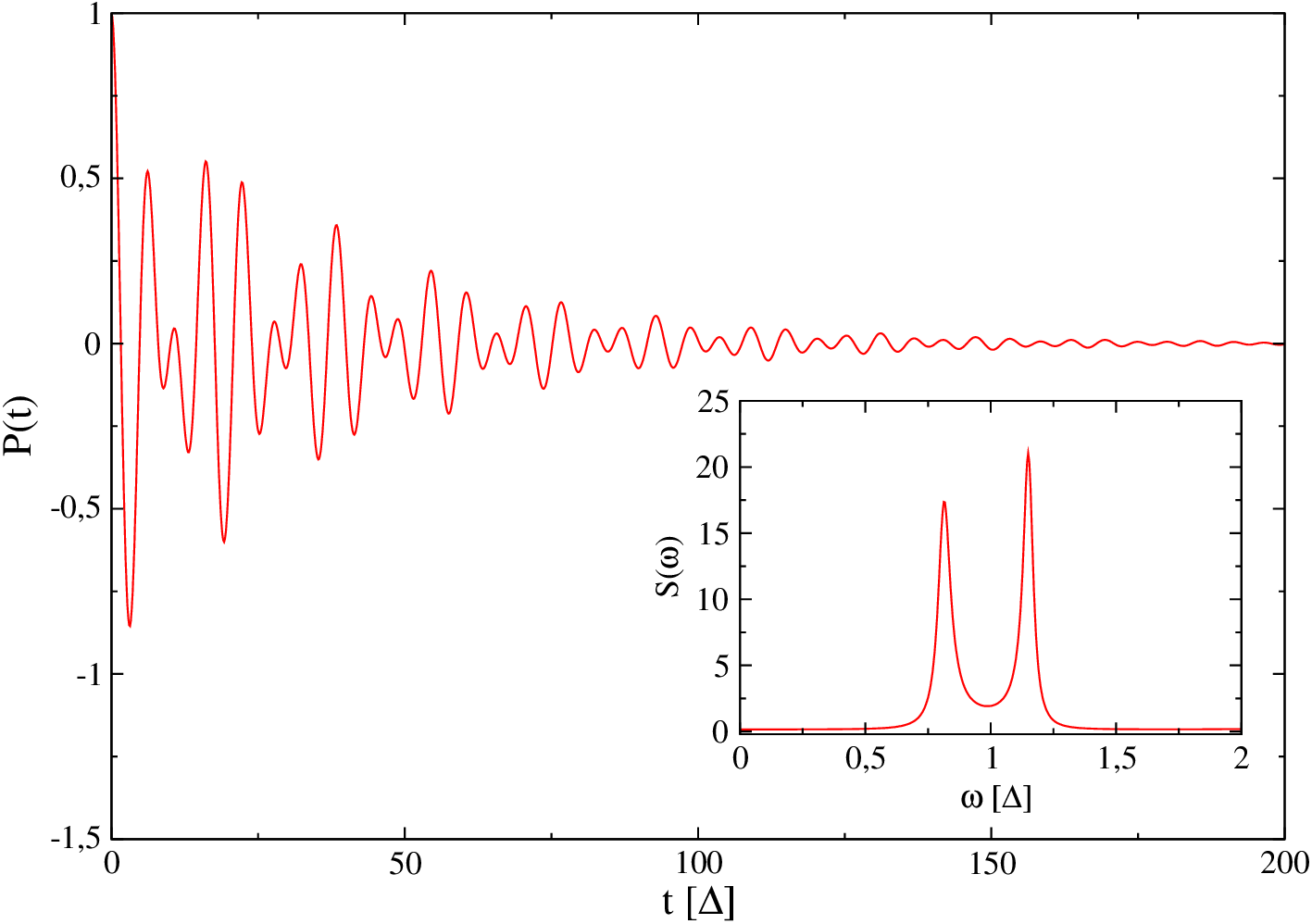}
\caption{Time evolution of the population difference $P(t)$ of a symmetric TSS in the resonant case $\Omega=\Delta$. The parameters are:  $\Gamma=0.097$, $\alpha=4\, 10^{-3}$, $(g=0.18)$, $T=0.1$ 
(all quantities are expressed in units of $\Delta$). In this range of parameters, one clearly sees that the dynamics is dominated by \textit{two} frequencies. Inset: Fourier Transform of $P(t)$. One clearly sees the appearance of two
peaks, centered symmetrically around $\Omega\pm g \approx 1\pm 0.18$.}
\label{pniba}
\vspace{0.5cm}
\includegraphics[width=0.7\textwidth,bb=47 57 709 528]{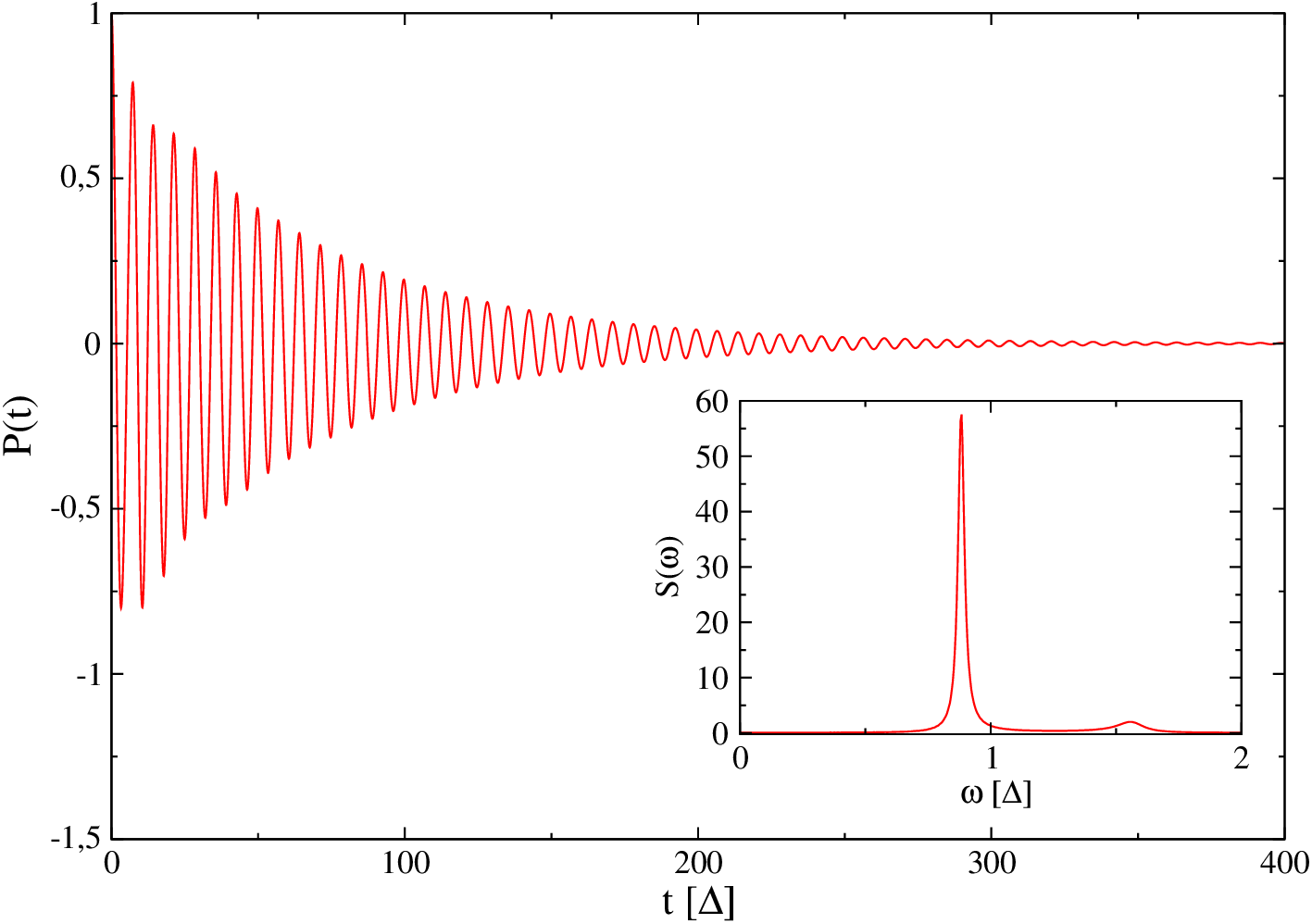}
\caption{Time evolution of the population difference $P(t)$ of a symmetric TSS in the case of a finite detuning $\Omega=1.5 \Delta$. The parameters are: $\Gamma=0.145$, $\alpha=5\, 10^{-3}$, $(g=0.3)$, $T=0.1$ 
(all quantities are expressed in units of $\Delta$). For a TSS being off-resonance with the HO, one notices that the relative magnitude of the two peaks become larger and larger, the higher the detuning is (see the Fourier Transform in the inset).  
} \label{pniba1.5}
\end{center}
\end{figure}

Within the NIBA \cite{Leggett87,Wei}, of the exact series expression for $\Ks(\lambda)$ 
only the first term of second order in the tunneling frequency $\Delta$ is retained.
This approximation has been commonly used over the whole range of temperatures and coupling strength to describe the dynamics of an unbiased TSS. 
In the undriven case, the model is justified for weak damping since the neglected correlations are of second order in the coupling $\alpha$, whereas for high temperature and/or large damping extra correlations are exponentially suppressed. 
The kernel has the very simple form 
\begin{align} 
\Ks(t)&=\Delta^{2} e^{-Q'(t)} 
\cos{\left(Q'' \left( t \right)\right)},
\end{align}
or in the Laplace space 
\begin{align} 
\Ks(\lambda)&=\Delta^{2} \intg{0}{\infty}\tau\, e^{-\lambda \tau} e^{-Q'(\tau)} 
\cos{\left(Q'' \left( \tau \right)\right)}\label{knibas},
\end{align}
where the bath correlation functions $Q'(\tau)$ and $Q''(\tau)$ have been introduced in Eq.\ \eqref{Q}. 
Typical results for $P(t)$ obtained from the numerical integration of the NIBA master equation for the resonant case $\Omega=\Delta$ and at finite detuning $\Omega=1.5 \Delta$ are shown in Figs.\ \ref{pniba} and \ref{pniba1.5}, respectively. In the resonant case 
 $P(t)$ exhibits a very pronounced beating pattern. The analysis of the corresponding spectrum 
\begin{equation}
S(\omega)\equiv 2\intg{0}{\infty}t \cos{(\omega t)}\,P(t) 
\end{equation}
for the parameters choice of \fig{pniba} (resonant case) clearly reveals the presence of two frequencies, which lie around $\Omega_\pm \approx \Omega \pm g$, where $g$ is the coupling strength in the TSS+HO model, as discussed above.
This is in agreement with predictions of a three-level system Bloch-Redfield analysis (with
second-order perturbation theory in $g$) as well as with exact results obtained within the numerical real-time path-integral approach QUAPI \cite{Tho03}. 
The Fourier spectrum for the detuned case in \fig{pniba1.5} shows a more pronounced oscillation frequency, the relative magnitude of the two peaks becoming larger and larger, the higher the detuning is. As one raises the coupling strength $\alpha$ between TSS and effective environment, multiple resonances appear, due to the fact that higher orders in Bessel functions contribute to the dynamics (see discussion below in the next Section).
 This beating pattern clearly originates from the peaked nature of the environmental spectrum and it is thus absent for the more frequently investigated cases of unstructured environments \cite{Leggett87,Wei}, i.e. $G(\omega) \propto \omega^s e^{-\omega/\omega_c}$, $s>0$. Starting point is \eq{P} and its related pole equation \eqref{peq}. 
The nature of the beatings as well as an analytical approximation to $P(t)$ are discussed in the following Section.

\section{Weak-Damping Approximation (WDA) for a symmetric TSS} \label{swca}

For a symmetric TSS, 
NIBA is expected to be justified in the whole regime of parameters. In particular we have seen, cfr.\ \fig{pniba}, that in the regime of resonance and strong coupling, i.e. $g \gg \Gamma$, it predicts
the two oscillation frequencies already found in Ref. \cite{Tho03} within a three-level approximation. However, the analysis in Ref. \cite{Tho03} was restricted to the case $\Delta=\Omega \gg k_{\rm B} T$. 
In the following, we shall derive an analytical expression for $P(t)$ valid for \textit{arbitrary} detuning $|\Delta-\Omega|\ne 0$ and low-to-moderate temperatures $k_{\rm B}T \lesssim \hbar\Omega$.
The key idea is that, since we are looking to a sharply peaked spectral density, i.e. $\kappa = \Gamma/2\pi\Omega \ll 1$, an expansion of the NIBA symmetric kernel \eqref{knibas} up to first order in $\kappa$ is justified. 

Since the bath-correlation functions $Q'$ and $Q''$ (\eq{Q''niba}) depend in a nontrivial way on $\kappa$, this requires some attention. 
In the end we obtain 
\begin{subequations} \label{qwca}
\begin{align}
Q'(\tau)&=\overbrace{Y\left(\cos{\Omega \tau}-1\right)}^{Q'_0(\tau)}+\overbrace{A \tau \cos{\Omega \tau}+B \tau+C \sin{\Omega \tau}}^{Q'_1(\tau)} + \cal{O}[\mbox{$\kappa^2$}]\,,\\
Q''(\tau)&=\underbrace{W\sin{\Omega
\tau}}_{Q''_0(\tau)}+\underbrace{V\left(1-\cos{\Omega
\tau}-\frac{\Omega}{2}\tau \sin{\Omega
\tau}\right)}_{Q''_1(\tau)}+\cal{O}[\mbox{$\kappa^2$}],
\end{align}
\end{subequations}
with the zero-order terms
\begin{align}
Y=-\frac{4 g^2}{\Omega^2}\: \frac{\sinh{\beta\Omega}}{\cosh{\beta\Omega}-1}, \qquad W=\frac{4 g^2}{\Omega^2},
\end{align}
and first-order terms
\begin{align}
A&=-\Gamma \frac{Y}{2},\\
B&=\Gamma \frac{8 g^2}{\Omega^3 \hbar \beta}, \\
\label{C}
C&=-\Gamma\frac{2 g^2}{\Omega^3}\: \frac{\beta\Omega+2 \sinh{\beta\Omega}}{\cosh{\beta\Omega}-1}, \\
V&=\Gamma \frac{4 g^2}{\Omega^3}.
\end{align}

Notice that the contribution coming from the Matsubara frequencies \eqref{Qmats} has been neglected, affecting only the short-time dynamics.
We will here discuss first the simpler undamped case ($\kappa=0$) and later perform the weak-damping approximation on the NIBA kernels.

\subsection{Undamped case ($\kappa=0$)}

In this Subsection we discuss the case of a TSS coupled with an
undamped HO initially prepared in a thermal equilibrium state.
The pole equation reads now
\begin{equation} \label{peqp}
\lambda_p+\Ks_0(\lambda_p)=0\,,
\end{equation}
with 

\begin{gather}  \label{ks0}
\hspace{-0.4cm} 
\Ks_0(\lambda)=\Delta^2 \intg{0}{\infty}\tau
e^{-\lambda\tau}\,e^{-Q'_0(\tau)} \cos{(Q''_0(\tau))}, \\
\end{gather}
%
where we denoted with $\lambda_p$ the solution of the undamped pole equation. 
Notice that $\Ks_0(\lambda)$ has the same expression as in
\eq{knibas}, if one replaces $Q'$ and $Q''$ with $Q'_0$ and
$Q''_0$, respectively. 
In order to evaluate \eq{ks0} analytically, we replace $\cos{\left(Q''_0 \left( \tau \right)\right)}$ with
$\mathfrak{Re}\{\exp{\left(\ii Q''_0 \left( \tau
\right)\right)}\}$ and we perform the Jacobi-Anger expansion \cite{Gradshteyn65} 
\begin{align} \label{bessel}
e^{\ii z \cos{y}} 
         &\equiv J_0(z)+2\sum_{n=1}^{+\infty} \ii^n J_n(z) \cos{(ny)} \,,
\end{align}
%
where $J_n(z)$ are Bessel functions of a complex argument. 
These
expansions are valid only in the case in which $z$ is independent
of $y$, which is the case for a TSS coupled to an undamped HO. 
We also make use of Graf's Addition Theorem
\begin{equation}
\sum_{k=-\infty}^\infty
J_{n+k}(u) J_k(v) \atp{\cos}{\sin} (k\alpha) = J_n(w) \atp{\cos}{\sin} (n\chi) ,
\end{equation}
where
\begin{equation}
w=\sqrt{u^2+v^2-2\,u\,v\,\cos{\alpha}}
\end{equation}
and
\begin{subnumcases}{}
&\hspace{-0.5cm}$u-v\cos{\alpha}= w \cos{\chi}$,\\
&\hspace{-0.5cm}$v\sin{\alpha}= w \sin{\chi}$.
\end{subnumcases}

We finally obtain 
\begin{equation} \label{k0J1}
\Ks_0(\lambda)=\Delta^{2} e^Y \intg{0}{\infty}\tau e^{-\lambda\tau}  \mathfrak{Re}\left[J_0(u_0)+2\sum_{n=1}^{+\infty} \ii^n J_n(u_0) \cos{[n(\Omega t-x)]}\right] \,
\end{equation}
where (see \ref{xapp})
\begin{subequations}
\begin{align} \label{u1}
u_0&=\frac{\ii Y}{\cos x} = \ii \sqrt{Y^2-W^2}
=\ii \frac{4 g^2}{\Omega^2} \dfrac{1}{\sinh{\left(\frac{\beta\Omega}{2}\right)}},\\
x&=\pi+\ii \frac{\beta\Omega}{2} \quad \left(\tan x=-\ii
\frac{W}{Y}\right). \label{x-u1}
\end{align}
\end{subequations}

After expanding the cosine which appears in the \eq{k0J1} and
after noticing that $J_0(u_0)$ and $\ii^n J_n(u_0)$ are always
real, the espression for the symmetric kernel in the undamped case finally reads
\begin{equation} \label{k0J2}
\Ks_0(\lambda)=\Delta^{2} e^Y \intg{0} {\infty}\tau e^{-\lambda\tau}  \left[J_0(u_0) +2\sum_{n=1}^{+\infty} (-\ii)^n J_n(u_0) \cos{(n\Omega\tau)}\cosh{\left(n\frac{\beta\Omega}{2}\right)}\right]  .
\end{equation}

It is useful to introduce here some amplitudes, in order to
enhance the readability of the kernel. We therefore define
\begin{subequations} \label{deltan0}
\begin{align} \label{deltanc0}
\Delta_{n_{(c)}}&\equiv\Delta
e^{Y/2}\sqrt{(2-\delta_{n,0})\,(-\ii)^n J_n(u_0)
\cosh{\left(n\frac{\beta\Omega}{2}\right)}} \,,
\end{align}
\end{subequations}
such that we can rewrite \eq{k0J2} in the very compact form
\begin{equation} \label{kwcaJ00}
\Ks_0(\lambda)=\sum_{n=0}^{+\infty} \Delta_{n_{(c)}}^{2}
\intg{0}{\infty}\tau e^{-\lambda \tau} \cos{(n\Omega\tau)}.
\end{equation}



The population difference $P_0(\lambda)$ in the undamped case becomes
\begin{align} \label{kwcaJ0}
 P_0(\lambda)&=\dfrac{1}{\lambda + \Ks_0(\lambda)}\\
&=\dfrac{1}{\lambda\left[1 + \sum_{n=0}^{+\infty} \Delta_{n_{(c)}}^{2} \dfrac{1}{\lambda^2+n^2\Omega^2}\right]}\\
&= \dfrac{\lambda^2\prod_{n=1}^\infty
(\lambda^2+n^2\Omega^2)}{\lambda \left[ \prod_{n=0}^\infty
(\lambda^2+n^2\Omega^2) +\sum_{m=0}^{+\infty}
\Delta_{m_{(c)}}^{2} \underset{n\neq
m}{\prod_{n=0}^{+\infty}}(\lambda^2+n^2\Omega^2) \right]} \,,
\label{P0}
\end{align}
and it is clear that the pole in $\lambda=0$ is not a physical
one, since $P_0(\lambda=0)$ vanishes. This means that the dissipation-free
($\kappa=0$) pole equation reads
\begin{equation} \label{peqpf2}
 \lambda_p +\Ks_0(\lambda_p)=0 \quad \to \quad  \prod_{n=0}^\infty (\lambda_p^2+n^2\Omega^2) +\sum_{m=0}^{+\infty}  \Delta_{m_{(c)}}^{2} \underset{n\neq m}{\prod_{n=0}^{+\infty}}(\lambda_p^2+n^2\Omega^2)  =0.
\end{equation}


\subsection{The weak-damping population difference $P(t)$}

The weak-damping kernel $\Ks_{WDA}(\lambda)$ is obtained from \eq{knibas} by retaining only terms up to first order in the linearized in $\kappa$ bath correlation functions $Q'_1$ and $Q''_1$. It reads 
\begin{gather}  \label{kswca0}
\Ks_{WDA}(\lambda) = \Delta^{2} \!\! \intg{0}{\infty}\tau e^{-\lambda\tau} \,e^{-Q'_0(\tau)} \left\{ \cos{(Q''_0(\tau))} \left[ 1-Q'_1(\tau) \right] -\sin{(Q''_0(\tau))} Q''_1(\tau) \right\}\!.
\end{gather}
The WDA kernel will be used in \eq{peq} to solve the pole equation and finally obtain $P_{WDA}(t)$. 
Consistent with the previous prescription $\kappa\ll 1$, we can expand the solutions $\lambda^*$ of the
pole equation around the solutions $\lambda_p$ of the
non-interacting pole equation up to first order in $\kappa$
 In other terms
\begin{equation} \label{lambdap}
\lambda^*=\lambda_p-\kappa\gamma_p+\ii\kappa\varphi,
\end{equation}
where $\lambda_p$ satisfies the undamped pole equation \eqref{peqpf2}. 
%
%
%
%
By inserting \eq{qwca} and \eqref{lambdap} in \eq{kswca0}, one
finds the following expressions for the kernels (to first order in
$\kappa$) at the poles:
\begin{align}  \label{kswca}
 \begin{split}
\Ks_{WDA}(\lambda^*)&=\Delta^{2} \intg{0}{\infty}\tau e^{-\lambda_p\tau} \,e^{-Q'_0(\tau)} \\
 &\times \left\{ \cos{(Q''_0(\tau))} \left[ 1+\kappa\gamma_p\tau-\ii\kappa\varphi\tau-Q'_1(\tau) \right] -\sin{(Q''_0(\tau))} Q''_1(\tau) \right\}+\cal{O}[\mbox{$\kappa^2$}].
\end{split}
\end{align}

According to \eq{kswca}, the pole equation \eqref{peq} now reads
\begin{align}
 \begin{split}
-\kappa\gamma_p+\ii\kappa\varphi &+\Delta^{2} \intg{0}{\infty}\tau
e^{-\lambda_p\tau} \,e^{-Q'_0(\tau)} \\
&\times \left\{ \cos{(Q''_0(\tau))}
\left[ \kappa\gamma_p\tau-\ii\kappa\varphi\tau-Q'_1(\tau)
\right]-\sin{(Q''_0(\tau))} Q''_1(\tau) \right\}=0\,,
\end{split}
\end{align}
where we used the pole equation for the undamped case \eqref{peqp}. 
%
After isolating the real and the imaginary terms from the above
equation \footnote{Note that the \lt\ of an odd function of $\tau$ is
even in $\lambda$ and vice versa. In this case, the integrand is
odd in $\tau$, thus the corresponding \lt\ is even in $\lambda$:
For pure-imaginary values of $\lambda$, the result of the integral
is real.}, we find
\begin{subequations}
\begin{align}
\begin{split}
-\kappa\gamma_p &\left[ 1-\Delta^{2} \intg{0}{\infty}\tau
e^{-\lambda_p\tau} \,e^{-Q'_0(\tau)} \cos{(Q''_0(\tau))}
\,\tau\right] \\
&= \Delta^{2} \intg{0}{\infty}\tau e^{-\lambda_p\tau} \,e^{-Q'_0(\tau)} \left[ \cos{(Q''_0(\tau))}\: Q'_1(\tau)+\sin{(Q''_0(\tau))} Q''_1(\tau) \right],
\end{split}\\
\ii\kappa\varphi &\left[ 1-\Delta^{2} \intg{0}{\infty}\tau
e^{-\lambda_p\tau} \,e^{-Q'_0(\tau)} \cos{(Q''_0(\tau))}
\,\tau\right] =0.
\end{align}
\end{subequations}
If the term between brackets is different from zero, one easily
gets $\varphi=0$ and, after some rearrangements, 
%
\begin{align}\label{gm}
\gamma_p= -\dfrac{\Delta}{\kappa}^{2}
\dfrac{\intg{0}{\infty}\tau e^{-\lambda_p\tau} \,e^{-Q'_0(\tau)}
\left[ \cos{(Q''_0(\tau))}\: Q'_1(\tau) +\sin{(Q''_0(\tau))}
Q''_1(\tau) \right]}
{\left[ 1+\dfrac{\partial}{\partial\lambda}\Ks_0(\lambda)\right]_{\big|_{\lambda=\lambda_p} }} \,.
\end{align}

Once we have obtained the expression for the decay rates $\gamma_p$
corresponding to each pole $\lambda_p$, we have all ingredients to
get the population difference $P(t)$ with the help of the Residue Theorem.
In fact, it holds 
\begin{align}
 P & (t) \equiv \sum_{\rm Res} e^{\lambda t} P(\lambda) \\
\begin{split}
&= \sum_{\lambda_p} e^{\lambda_p t} e^{-\kappa \gamma_p(\lambda_p) t} {\displaystyle \lim_{\lambda=\lambda_0-\kappa \gamma \to \lambda_p-\kappa \gamma_p}} [\lambda - (\lambda_p-\kappa \gamma_p)] \dfrac{1}{\lambda+\Ks_{WDA}(\lambda)}\,,
\end{split}
\end{align}
as follows from \eq{kswca}. Notice that here we split the damping-dependent and the damping-indipendent contributions as $\lambda=\lambda_0-\kappa \gamma$. 
The limit ${\displaystyle
\lim_{\lambda=\lambda_0-\kappa \gamma \to \lambda_p-\kappa
\gamma_p}}$ can also be rewritten as 
${\displaystyle
\lim_{\lambda_0 \to \lambda_p}  \lim_{\gamma \to
\gamma_p(\lambda_0)}}$. 
Hence, performing first the limit over
the decay rate $\gamma$, we find
\begin{align} \label{Pprima}
\begin{split}
 P(t)&=\sum_{\lambda_p} e^{\lambda_p t} e^{-\kappa \gamma_p t} {\displaystyle \lim_{\lambda_0 \to \lambda_p}} (\lambda_0-\lambda_p) 
\left[ \lambda_0 +\Delta^2 \intg{0}{\infty}\tau e^{-\lambda_0\tau} \,e^{-Q'_0(\tau)} 
\cos{(Q''_0(\tau))} \right. \\
&\left.  - \kappa \left( 1-\Delta^2 \intg{0}{\infty}\tau e^{-\lambda_0\tau} \,e^{-Q'_0(\tau)} 
\cos{(Q''_0(\tau))} \tau \right) \left(\gamma_p(\lambda_0)-\gamma_p(\lambda_p) \right) \right]^{-1}
\end{split}
\end{align}
and performing the limit over $\lambda_0$,  $P(t)$ finally reads 
\begin{align} \label{Pfin}
 P(t)
&=\sum_{\lambda_p} e^{\lambda_p t} e^{-\kappa \gamma_p t}
{\displaystyle \lim_{\lambda_0 \to \lambda_p}}
(\lambda_0-\lambda_p) P_0(\lambda_0) \,,
\end{align}
as it follows from Eq. \eqref{kwcaJ0}. 
We stress that the dynamics in the (weakly) damped case is essentially
determined by the correponding undamped dynamics, the damping
being only responsible for pole-dependent exponentially decaying factors.

\subsection{Series expression for the weakly-damped symmetric kernel and decay rate}

Now we want to find a compact analytical form for the kernel $\Ks_{WDA}(\lambda)$ and the decay rate $\gamma_p$ 
\eqref{kswca0} and \eqref{gm}, respectively. To this extent, let us start from
the kernel $\Ks_{WDA}(\lambda)$, the generalization to the decay rate being 
straightforward.

As in the undamped case, in \eq{kswca0} we replace $\cos{\left(Q''_0 \left( \tau \right)\right)}$ with
$\mathfrak{Re}\{\exp{\left(\ii Q''_0 \left( \tau
\right)\right)}\}$ and $\sin{\left(Q''_0 \left( \tau
\right)\right)}$ with $\mathfrak{Im}\{\exp{\left(\ii Q''_0 \left(
\tau \right)\right)}\}$. Analogously to the procedure followed for the undamped kernel \eqref{kwcaJ00}, by using the Jacobi-Anger expansion \eqref{bessel} we obtain 
\begin{equation} \label{kwcaJc}
\Ks_{WDA}(\lambda)=\sum_{n=0}^{+\infty} \intg{0}{\infty}\tau e^{-\lambda
\tau} \left\{\Delta_{n_{(c)}}^{2} \cos{(n\Omega\tau)} \left[
1-Q'_1(\tau) \right] +\Delta_{n_{(s)}}^{2} \sin{(n\Omega\tau)}
Q''_1(\tau) \right\},
\end{equation}
where $u_0$ and $x$ are given by \eq{u1} and \eq{x-u1}. The dressed tunneling elements $\Delta_{n_{(c)}}$ have been already defined in \eq{deltanc0} and 
\begin{align} \label{deltans}
\Delta_{n_{(s)}}&\equiv\Delta
e^{Y/2}\sqrt{(2-\delta_{n,0})\,(-\ii)^n J_n(u_0)
\sinh{\left(n\frac{\beta\Omega}{2}\right)}}.
\end{align}
The expression for $P(\lambda)$ follows from \eqref{kwcaJc} (see the discussion in the Section \ref{Pan} below).
Along similar lines, the decay rate $\gamma_p$, cf. \eq{gm}, may also be written
as
%
%
%
%
%
\begin{equation} \label{ratesf}
\gamma_p= \dfrac{1}{\kappa} \dfrac{\sum_{n=0}^{+\infty}
\intg{0}{\infty}\tau e^{-\lambda_p\tau} \left[
\Delta_{n_{(c)}}^{2} \cos{(n\Omega\tau)}\: Q'_1(\tau) +
\Delta_{n_{(s)}}^{2} \sin{(n\Omega\tau)}\: Q''_1(\tau) \right]
}{\sum_{n=0}^{+\infty} \Delta_{n_{(c)}}^{2}
\dfrac{2\lambda_p^2}{(\lambda_p^2+n^2\Omega^2)^2}}.
\end{equation}

\subsection{The case $n=0,\: n=1$}

In order to obtain a more useful analytical expression for $P(t)$, 
we notice that the lowest orders in $n$ give the largest
contribution to the sum in \eq{kwcaJc}, because the amplitudes
$\Delta_n^2$ depend on Bessel functions $J_n(x)$, which roughly
behave as $x^n$ as soon as the argument becomes small. Since we investigate a regime of temperatures generally smaller than $\Omega$, i.e. $\beta \hbar\Omega/2 \gtrsim 1$, and of coupling such that in general $g \lesssim \Omega$, then the quantity $u_0$ is smaller than one. 
Hence, we just restrict our analysis to the
orders $n=0, n=1$ in Eqs. \eqref{kwcaJc} and \eqref{ratesf}. The
undamped pole equation \eq{peqpf2} therefore becomes (we identify
here $\Delta_0$ with $\Delta_{0_{(c)}}$ for the sake of clarity)
\begin{equation}  \label{pole-eq-non}
(\lambda_p^2+\Omega^2) (\lambda_p^2+\Delta_0^{2})
+\lambda_p^2\Delta_{1_{(c)}}^{2}=0\,,
\end{equation}
yielding
\begin{equation}
\label{poli}
\lambda_{p}^2=-\frac{\Delta_0^2+\Delta_{1_{(c)}}^{2}+\Omega^2}{2}
\pm\sqrt{\left(\frac{\Delta_0^2-\Omega^2}{2}\right)^2+\dfrac{\Delta_{1_{(c)}}^{2}}{2}
\left(\Delta_0^2+\dfrac{\Delta_{1_{(c)}}^{2}}{2}+\Omega^2\right)} \equiv \lambda_\pm^2\,.
\end{equation}
We notice that only terms quadratic in $\lambda_p$ appear in the formal expression of the decay rate (cf. \eq{gammaf}). Hence, it is enough to express the poles as in \eq{poli}.


Given the poles in the undamped case, we can substitute each of them in the Eq. \eqref{ratesf} for $\gamma_p$ with sum restricted to $n=0, n=1$. We will refer to them as $\gamma_{\pm}=\gamma(\lambda_{\pm})$, 
%
%
the explicit form of the decay rate being given in \ref{gammap}. 

\subsection{Analytical expression for $P(t)$} \label{Pan}

In order to obtain the analytical expression for $P(t)$ in the symmetric case, let us start again from \eq{Pfin}. By summing up all residues contributions, we end up with 
\begin{align} 
\begin{split}
P(t)&=
e^{-\kappa\gamma_- t}\: \frac{\lambda_-^2+\Omega^2}{\lambda_-^2+\lambda_+^2}\cos{\Omega_- t}+
e^{-\kappa\gamma_+ t}\: \frac{\lambda_+^2+\Omega^2}{\lambda_+^2+\lambda_-^2}\cos{\Omega_+ t}\\
& \hspace{1.5cm}-e^{-\kappa\gamma_- t}\:
\frac{\kappa\gamma_-}{\Omega_-}\:
\frac{\lambda_-^2+\Omega^2}{\lambda_-^2-\lambda_+^2}\sin{\Omega_-
t}
-e^{-\kappa\gamma_+ t}\: \frac{\kappa\gamma_+}{\Omega_+}\: \frac{\lambda_+^2+\Omega^2}{\lambda_+^2-\lambda_-^2}\sin{\Omega_+ t}, \end{split}  \label{Pt}
\end{align}
where $\Omega_\pm \equiv -\ii \lambda_\pm$, as follows from \eq{poli}.

\begin{figure}[htb!]
\begin{center}
\includegraphics[width=0.7\textwidth,bb=34 48 718 527,clip=true]{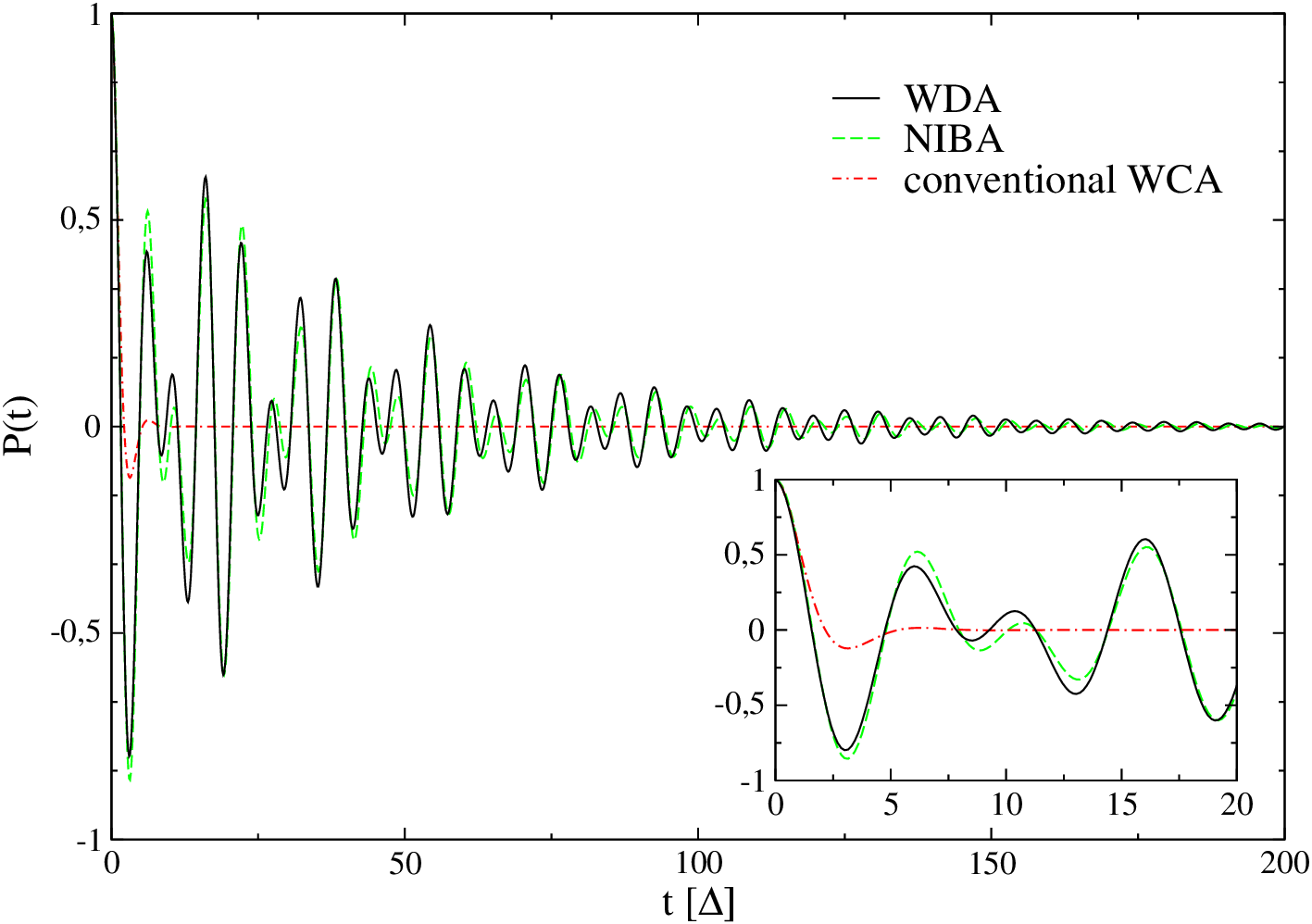}
\vspace{-0.1cm}\caption{Time evolution of $P(t)$ within the NIBA as well as the analytical WDA.  
The parameters are as in \fig{pniba}, namely 
$\Omega=\Delta$, 
$\Gamma=0.097$, $\alpha=4\, 10^{-3}$, 
$(g=0.18)$, $T=0.1$ (in units of $\Delta$). Notice the perfect 
agreement between the numerical NIBA and the analytical WDA. A perturbative approach in $G_{\rm eff}(\omega)$, denoted here as ``conventional WCA'' (see \eqref{wcaold}), completely fails because it does not even account for the two main oscillation frequencies. In the inset the short-time dynamics is magnified.}
\label{plot}
\vspace{0.5cm}
\includegraphics[width=0.7\textwidth,bb=43 48 709 528,clip=true]{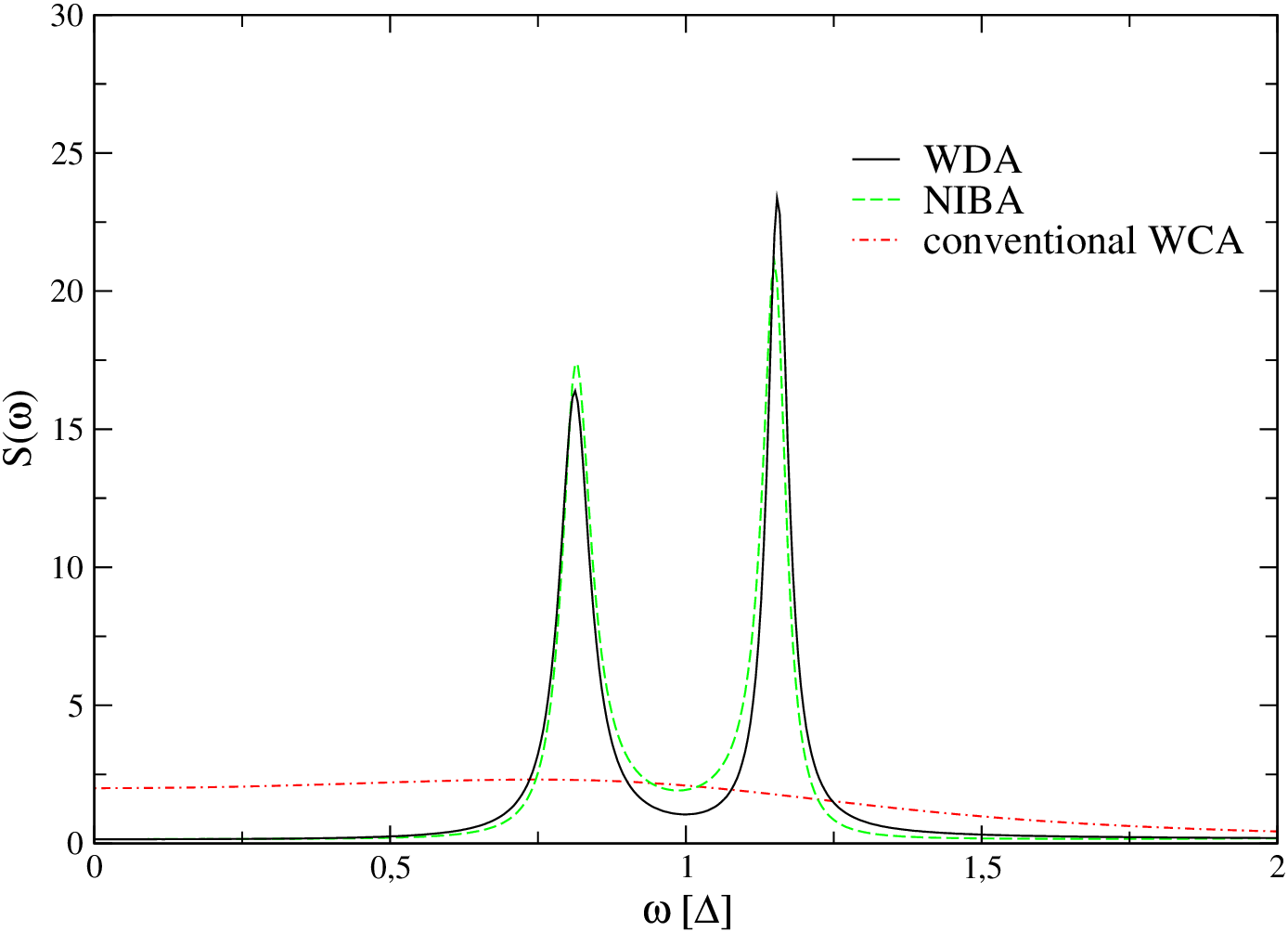}
\vspace{-0.1cm}\caption{Spectral function of $P(t)$, corresponding to the same regime as in the previuos case.
One clearly sees that the Fourier Transforms of NIBA and WDA exhibit a double peak structure. In contrast, the WCA predicts a single broadened oscillation peak.
}
\label{ft}
\end{center}
\end{figure}
Notice that the expression for $P(t)$ is invariant upon exchanging the frequencies $\Omega_- \rightarrow \Omega_+$.

 The analytical formula \eq{Pt} for $P(t)$ is compared in \fig{plot} with the exact numerical NIBA and the conventional weak-coupling approximation (WCA), obtained by a linear expansion of the bath correlation function for the coupling strength $\alpha$ \cite{Wei}.
There, the analytical form for the probability difference reads in the symmetric case 
\begin{equation} \label{wcaold}
\begin{split}
 P(t) &= \left\{  \cos{\Delta t}+ \dfrac{\gamma_\varphi}{\Delta}
  \sin{\Delta t} \right\} e^{-\gamma_\varphi t},
\end{split}
\end{equation}
where $\gamma_\phi=\frac{\pi}{4} S(\Delta)$ 
is the dephasing rate. Moreover, $S(\omega) \equiv G_{\rm eff}(\omega) \coth{\left(\frac{\hbar \omega}{2 k_{\rm B} T}\right)}$ is a spectral contribution which represents emission and absorption of a single photon.
The choice of parameters in Figs.\ \ref{plot} and \ref{ft} is the same as in \fig{pniba} and as Ref.\ \cite{Tho03}, under the resonance condition $\Omega=\Delta$. 
\begin{figure}[htb!]
\begin{center}
\includegraphics[width=0.7\textwidth,bb=34 48 718 527,clip=true]{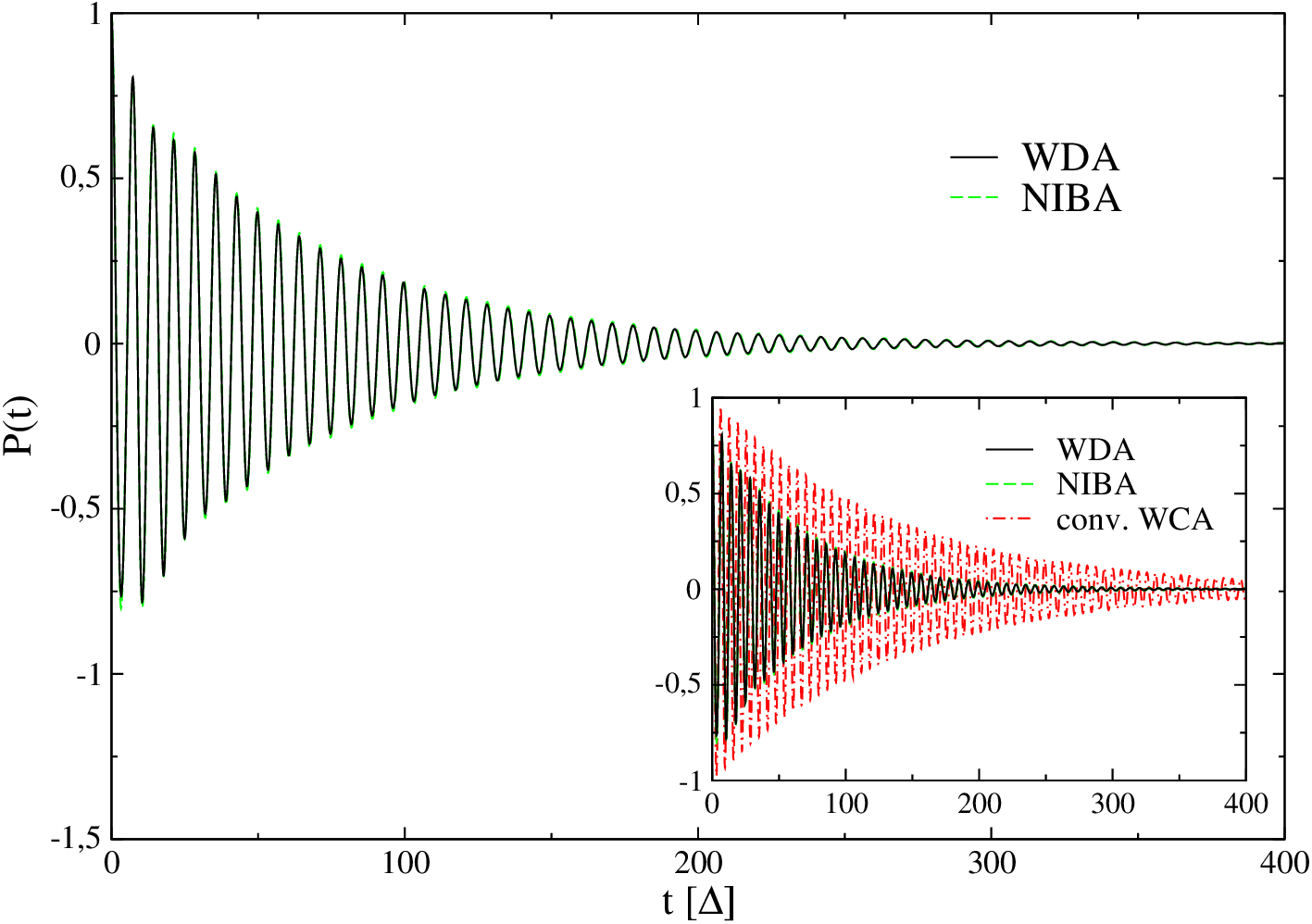}
\vspace{-0.1cm}\caption{Time evolution of $P(t)$ at finite detuning within the NIBA and the analytical WDA. 
The parameters are as in \fig{pniba1.5}, namely 
$\Omega=1.5\Delta$, 
$\Gamma=0.145$, $\alpha=5 \, 10^{-3}$, 
$(g=0.3)$, $T=0.1$ (in units of $\Delta$). Again, the
agreement between the numerical NIBA and the analytical WDA is striking. The conventional WCA \eqref{wcaold} (see inset)
also in this case, as expected, fails in describing the correct dynamics.}
\label{pt1.5}
\vspace{0.5cm}
\includegraphics[width=0.7\textwidth,bb=43 48 709 528,clip=true]{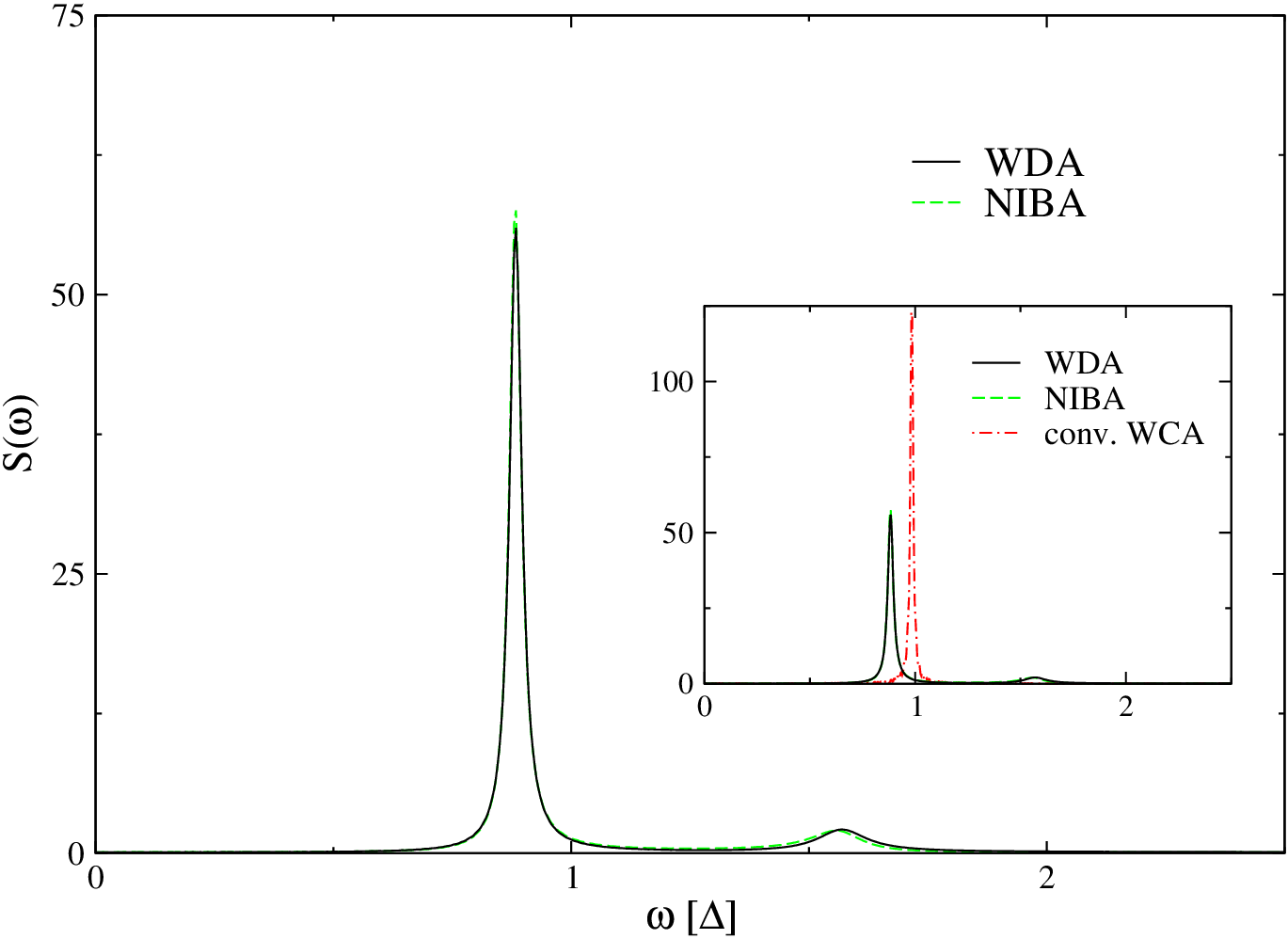}
\vspace{-0.1cm}\caption{Spectral function of $P(t)$ for the NIBA and the analytical WDA (same parameters as in the previous \fig{pt1.5}).
The oscillation frequencies of WDA and NIBA coincide, whereas in the inset one can at first glance see the appearance of a single peak concerning the Fourier Transform of the conventional WCA.
}
\label{ft1.5}
\end{center}
\end{figure}
One can notice a very good agreement between NIBA and the analytical WDA, whereas \eq{wcaold} completely fails in describing the  oscillatory behaviour of $P(t)$. 
In \fig{ft} the corresponding Fourier Transform of the probability difference is showed. There, one can see the missing oscillation frequency of the conventional WCA given by \eq{wcaold} and the excellent agreement between the numerical NIBA and our analytical solution WDA.

Finally, in \fig{pt1.5} we show a comparison among the WDA and the NIBA in presence of finite detuning $|\Delta-\Omega|=0.5$ for a higher coupling strength between qubit and HO ($g=0.3$), keeping the coupling between detector and environment constant. Also in this case, the weak-damping approximation fully agrees with the numerical solution of the NIBA. In the inset one can also notice the disagreement of both models with the conventional WCA, characterized by a single oscillation frequency at $\omega=\Delta$, see in particular \fig{ft1.5} which shows the Fourier Transform of $P(t)$.

\section{Conclusions}
In conclusion, we discussed the dynamics of a symmetric TSS interacting with an effective structured environment, modelling a qubit interacting with a readout dissipative detector. This case has not been so far deeply investigated within an analytical approach out of resonance, i.e. if the tunneling frequency $\Delta$ differs from the detection frequency $\Omega$.
We approximated an exact generalized master equation (GME) within a novel \textit{weak-damping approximation} (WDA) which, in contrast to ``conventional'' weak-coupling approaches \cite{Wei}, is able to correctly reproduce the dynamics, characterized by multiple oscillation frequencies.
The WDA approach is based on a first approximation of the kernel of the GME up to second order in the tunneling frequency, i.e. the Non-Interacting Blip Approximation (NIBA), which for a symmetric spin-boson model is valid over the \textit{whole} range of parameters. Then, in order to obtain an analytical form for the dynamical population difference $P(t)$, for small enough temperatures (i.e. $k_{\rm B}T \lesssim \hbar\Omega$) and small coupling strength between detector and environment, we approximated the NIBA kernels within a weak-damping approach, whose details are explained in Sec.\ \ref{swca}.
The agreement of our analytical solution for $P(t)$ valid at \textit{arbitrary} detuning $|\Delta-\Omega|\ne 0$ with the numerical NIBA is striking. The former one is able to reproduce the two oscillation frequencies which are related to the tunneling and the detection frequency, respectively, as predicted by \textit{ab-initio} numerical schemes like QUAPI \cite{Tho03}.
Our results are of interest for the understanding of dephasing in qubits strongly coupled to a broadened harmonic mode as e.g. flux qubits \cite{Chi03, Patrice04} or cavity QED qubits \cite{Wallr04}.

\section{ACKNOWLEDGMENTS} 
We acknowledge financial support under the DFG program SFB631.

\appendix
\section{Calculation of the parameter $x$} \label{xapp}

Here we are interested in finding out the correct value for $x$ which comes in the formulas after we make the expansion in Bessel functions. We refer here to the Eq. \eqref{kswca0}, which we rewrite for clarity.
\begin{gather}  
\Ks(\lambda)=\Delta^{2} \intg{0}{\infty}\tau e^{-\lambda\tau} \,e^{-Q'_0(\tau)} \left\{ \cos{(Q''_0(\tau))} \left[ 1-Q'_1(\tau) \right] -\sin{(Q''_0(\tau))} Q''_1(\tau) \right\}.
\end{gather}

Let us examine, to fix the ideas, the term ${\rm exp}\{-Q'_0(\tau)\} \cos{(Q''_0(\tau))}$:
\begin{align}
e^{-Q'_0(\tau)} \cos{(Q''_0(\tau))} &\equiv e^{Y} \mathfrak{Re}\left\{ e^{-Y \cos{\Omega\tau}} e^{+\ii W \sin{\Omega\tau}} \right\} \\
&= e^{Y} \mathfrak{Re}\left\{e^{\ii\left[\ii Y \cos{\Omega\tau}+ W \sin{\Omega\tau} \right]}\right\} \\
&= e^{Y} \mathfrak{Re}\left\{e^{\ii\sqrt{W^2-Y^2} \left[\dfrac{\ii Y}{\sqrt{W^2-Y^2}} \cos{\Omega\tau}+ \dfrac{W}{\sqrt{W^2-Y^2}} \sin{\Omega\tau} \right]}\right\}\,.
\end{align}
It could be now convenient to interprete
\begin{subnumcases} {\label{cossin}}
{}& \hspace{-0.5cm}$\cos{x} \equiv \dfrac{\ii Y}{\sqrt{W^2-Y^2}} = +\dfrac{Y}{\sqrt{Y^2-W^2}}$\\
{}& \hspace{-0.5cm}$\sin{x} \equiv \dfrac{-W}{\sqrt{W^2-Y^2}} = +\dfrac{\ii W}{\sqrt{Y^2-W^2}}$ \,,
\end{subnumcases}
 so that the exponent can be rewritten as
\begin{equation}
 e^{-Q'_0(\tau)} \cos{(Q''_0(\tau))} = e^{Y} \mathfrak{Re}\left\{ e^{\ii\sqrt{W^2-Y^2} \cos{(\Omega\tau+x)}}\right\} \,.
\end{equation}

At this point we can use the Jacobi-Anger expansion \eqref{bessel} to expand the exponent in series of Bessel functions.
We finally obtain 
\begin{equation} 
\begin{split}
\Ks(\lambda)=\Delta^{2} e^Y &\intg{0}{\infty}\tau e^{-\lambda\tau} \left\{ \left[J_0(u_0) +2\sum_{n=1}^{+\infty} (-\ii)^n J_n(u_0) \cos{(n\Omega\tau)}\cosh{\left(n\frac{\beta\Omega}{2}\right)}\right] \right. \\
& \times \left. \left[ 1-Q'_1(\tau) \right] +2\sum_{n=1}^{+\infty} (-\ii)^n J_n(u_0) \sin{(n\Omega\tau)}\sinh{\left(n\frac{\beta\Omega}{2}\right)} Q''_1(\tau) \right\} ,
\end{split}
\end{equation}
which coincides with \eq{kwcaJc}, once we introduce the amplitudes. Here,  $u_0$ is given by
\begin{equation}
 u_0 \equiv \sqrt{W^2-Y^2} = \ii\sqrt{Y^2-W^2} =\ii \frac{4 g^2}{\Omega^2} \dfrac{1}{\sinh{\left(\frac{\beta\Omega}{2}\right)}}\,,
\end{equation}
since $Y \equiv -W \coth{(\frac{\beta \Omega}{2})}$, hence $|Y| \ge |W|$ (note that $W>0$ and hence $Y<0$). One notices that the argument of the Bessel functions is small, whenever $\beta \hbar\Omega/2 \gtrsim 1$ and $g \lesssim \Omega$, i.e. in the regime we are interested in.

We would like now to obtain the exact value of $x$, which is easily performed. Let us start from Eq. \eqref{cossin} and let us rewrite the tangent as
\begin{equation}
 \tan{x}=\dfrac{+ \ii W}{Y}=-\ii \tanh{(\frac{\beta\Omega}{2})}=\tan{(-\ii\frac{\beta\Omega}{2})} \,.
\end{equation}

We assume $x$ to be complex, therefore we write it as $x=a+\ii b$. In general, it holds
\begin{eqnarray}
 \cos{(a+\ii b)}&=\cos{a} \cosh{b}-\ii \sin{a} \sinh{b} \label{cos}\\
 \sin{(a+\ii b)}&=\sin{a} \cosh{b}+\ii \cos{a} \sinh{b} \label{sin} \,.
\end{eqnarray}
From \eqref{cos}, in order to have $\cos{(a+\ii b)}=+Y/\sqrt{Y^2-W^2}$, namely a real number, it must be
\begin{equation}
 a=n \pi \,.
\end{equation}
From Eqs. \eqref{cos} and \eqref{sin}, we can write the tangent as
\begin{equation} \label{tan}
 \tan{(a+\ii b)}=\dfrac{\tan{a}+\ii \tanh{b}}{1-\ii \tan{a} \tanh{b}} \xrightarrow[a=n\pi]{} +\ii \tanh{b} \equiv +\ii \dfrac{W}{Y}\,,
\end{equation}
as we get by calculating the tangent from Eq. \eqref{cossin}.
Hence,
\begin{equation}
 \tanh{b}=\dfrac{W}{Y}.
\end{equation}
We can eventually write $x$ as
\begin{equation}
 x\equiv a+\ii b=n\pi+\ii\, {\rm arctanh}{\dfrac{W}{Y}}=n\pi-\ii \dfrac{\beta \Omega}{2}\,.
\end{equation}

In order to decide, whether to assume $n=0$ or $n=1$, one must look at the cosine or sine:
\begin{align}
 \cos{x} &\xrightarrow[a=n\pi]{{\rm Eq.} \eqref{cos}} (-1)^n \dfrac{1}{\sqrt{1-\tanh^2{b}}}= (-1)^n \dfrac{1}{\sqrt{1-\dfrac{W^2}{Y^2}}} =(-1)^n \dfrac{|Y|}{\sqrt{Y^2-W^2}} \\
&\equiv \dfrac{+Y}{\sqrt{Y^2-W^2}} <0 \hspace{1cm}\Longmapsto n=1 \\
\intertext{or, equivalently,}
 \sin{x} &\xrightarrow[a=n\pi]{{\rm Eq.} \eqref{sin}} \ii(-1)^n \dfrac{\tanh{b}}{\sqrt{1-\tanh^2{b}}}= \ii (-1)^n \dfrac{\dfrac{W}{Y}}{\sqrt{1-\dfrac{W^2}{Y^2}}} \\
&=\ii (-1)^n \dfrac{\dfrac{W}{Y}|Y|}{\sqrt{Y^2-W^2}} =-\ii (-1)^n \dfrac{W}{\sqrt{Y^2-W^2}}\equiv \dfrac{+\ii W}{\sqrt{Y^2-W^2}} \hspace{1cm}\Longmapsto n=1 \,.
\end{align}

\section{Explicit form for the decay rate $\gamma_p$} \label{gammap}

In this Appendix we wish to give the analytical result for the decay rate $\gamma_p(\lambda_p)$ as function of the solution $\lambda_p$ of the undamped pole equation \eq{pole-eq-non}:
\begin{equation} \label{gammaf}
\begin{split}
 \gamma_p(\lambda_p)=&\dfrac{1}{\kappa}\dfrac{1}{2\lambda_p^2\left[(\lambda_p^2+\Omega^2)^2+\Delta_{1_{(c)}}^2\Omega^2\right]}
\Bigg[
\lambda_p^2 \left(p+q\Delta_{1_{(c)}}^2\!\!+t\Delta_{1_{(s)}}^2\!\!+u\Delta_{1_{(c)}}^2\Delta_{1_{(s)}}^2\!\!+r\Delta_{1_{(c)}}^4\!\right) \\
&\qquad+\Omega^2 \left(s+w\Delta_{1_{(c)}}^2+t\Delta_{1_{(s)}}^2 \right)+\lambda_p^2\left(\Delta_{1_{(c)}}^2 g(\lambda_p)+\Delta_{1_{(s)}}^2 h(\lambda_p) \right) \Bigg] ,
\end{split}
\end{equation}
with
\begin{eqnarray}
p&\equiv\ &(2A-B)\Omega^2\Delta_0^2+(B-D)\Delta_0^4,\\
q&\equiv\ &\Delta_0^2(\dfrac{A}{2}+2B-D)+\Omega^2(2B-\dfrac{A}{2}),\\
t&\equiv\ &-\dfrac{V\Omega}{4} (\Omega^2+3\Delta_0^2),\\
u&\equiv\ &-3\dfrac{V\Omega}{4},\\
r&\equiv\ &\dfrac{A}{2}+B,\\
s&\equiv\ &-\Omega^2\Delta_0^2 B+\Delta_0^4 (B-D),\\
w&\equiv\ &\Delta_0^2(\dfrac{A}{2}+B)-\Omega^2 \dfrac{A}{2}
\end{eqnarray}
and
\begin{eqnarray}
g(\lambda_p)&\equiv& \frac{(\lambda_p^2+\Omega^2)^2}{\lambda_p^2+4\Omega^2}\left(C\Omega+\dfrac{A}{2} \frac{\lambda_p^2-4\Omega^2}{\lambda_p^2+4\Omega^2}\right), \\
h(\lambda_p)&\equiv& \frac{(\lambda_p^2+\Omega^2)^2}{\lambda_p^2+4\Omega^2}\dfrac{V\Omega}{4} \frac{3\lambda_p^2+20\Omega^2}{\lambda_p^2+4\Omega^2}.
\end{eqnarray}



 As already seen, the physical poles are 
$\lambda^2=-\lambda_{1,2}^2 \equiv \lambda_\pm^2$. 
Correspondingly, the decay rates $\gamma_{\pm}=\gamma(\lambda_{\pm})$ follow according to \eq{gammaf}.

\vspace{0.5cm}
\centerline{***}
\vspace{0.5cm}


\begin{thebibliography}{99}

\bibitem{Leggett87}A.J. Leggett {\em et al.}, Rev. Mod. Phys. {\bf 59}, 1 (1987).
\bibitem{Wei} U.\ Weiss, {\em Quantum Dissipative Systems\/}
(World Scientific, Singapore, 2nd ed., 1999).
\bibitem{Gri98} M.\ Grifoni and P.\ H\"{a}nggi, Phys.\ Rep.\ {\bf 304}, 229 (1998).
\bibitem{Makhlin01} Y.\ Makhlin, G.\ Sch\"{o}n and A.\ Shnirman, Rev. Mod. Phys. {\bf 73}, 357
(2001).
\bibitem{Wal00} C.\ van der Wal {\em et al.\/}, Science {\bf 290}, 773 (2000).
\bibitem{Chi03} I.\ Chiorescu {\em et al.\/}, Science {\bf 299}, 1869 (2003).
\bibitem{Patrice04} I. Chiorescu {\em et al.\/}, Nature {\bf 431}, 159 (2004).
\bibitem{Tian02} L.\ Tian, S.\ Lloyd, T.P.\ Orlando, Phys. Rev. B
{\bf 65}, 144516 (2002).
\bibitem{vdW} C.H.\ van der Wal, F.K.\ Wilhelm, C.J.P.M.\ Harmans and J.E.\ Mooij, Eur.\ Phys.\ J.\ B  {\bf 31}, 111 (2003).
\bibitem{Wallr04}A.\ Wallraff {\em et al.\/}, Nature {\bf 431}, 162 (2004).
%
\bibitem{Thorwart00}M.\ Thorwart {\em et al.\/},
 J.\ Mod.\ Opt.\ {\bf 47}, 2905 (2000).
 \bibitem{Smi03} A.\ Yu.\ Smirnov, Phys.\ Rev.\ B {\bf 67}, 155104 (2003).
\bibitem{Tho03} M.\ Thorwart, E.\ Paladino and M.\ Grifoni, Chem.\ Phys.\
{\bf 296}, 333 (2004).
\bibitem{Kle03} F.\ K.\ Wilhelm, S.\ Kleff and J.\ von Delft,
Chem.\ Phys.\ {\bf 296}, 345 (2004).
\bibitem{Goo03} M.\ C.\ Goorden and F.\ K.\ Wilhelm, Phys.\ Rev.\ B {\bf
68}, 012508 (2003).
\bibitem{Goo05} M.C.\ Goorden, M.\ Thorwart and M.\ Grifoni, Phys.\ Rev.\ Lett.\ {\bf 93}, 267005 (2004); Eur.\ Phys.\ J.\ B {\bf 45}, 405 (2005).
\bibitem{Garg85}A.\ Garg, J.\ N.\ Onuchic and V.\ Ambegaokar,
J.\ Chem.\ Phys.\  {\bf 83}, 4491 (1985).
\bibitem{Gradshteyn65} I.S.\ Gradshteyn and I.M.\ Ryzhik, {\em Tables of Integrals, Series and Products}
 (Academic Press, London, 1965).

\end{thebibliography}

\end{document}